\documentclass{ws-procs975x65}

\usepackage{psfrag}
\usepackage{mathrsfs} 
\usepackage{graphicx}
\usepackage{amsmath}
\usepackage{amssymb}
\usepackage{bm}
\usepackage{dsfont} 
\usepackage{pstricks}
\usepackage{slashed}
\usepackage{hyperref}

\newcommand{\ddx}{\textrm{d}^d x\:}
\newcommand{\Tr}{\text{Tr}\:}

\newcommand{\be}{\begin{equation}}
\newcommand{\ee}{\end{equation}}
\newcommand{\ba}{\begin{align}}
\newcommand{\ea}{\begin{align}}
\newcommand{\p}{\partial}
\newcommand{\sg}{\sqrt{g}}
\newcommand{\sgb}{\sqrt{\bar{g}}}
\newcommand{\gb}{\bar{g}}
\newcommand{\mn}{{\mu\nu}}
\newcommand{\rs}{{\rho\sigma}}
\newcommand{\Rk}{\mathcal{R}_k}
\newcommand{\ord}{\mathcal{O}}
\newcommand{\ZFk}{Z_{F,k}}
\newcommand{\sD}{\slashed{D}}

\newcommand{\RN}{R^{(0)}}

\newcommand{\mU}{\mathbf{U}} 
\newcommand{\bp}{\bm{\varphi}}
\newcommand{\cA}{\mathcal{A}}

\begin{document}

\markboth{A.~Nink and M.~Reuter}{On Quantum Gravity, Asymptotic Safety,
	and Paramagnetic Dominance}

\title{\uppercase{\ \\[-12.6mm] On Quantum Gravity, Asymptotic Safety,\\
	and Paramagnetic Dominance}\footnote{Plenary talk given by M.R.\ at the 13th Marcel
	Grossmann Meeting, Stockholm, 1--7 July 2012.}}

\author{\uppercase{Andreas Nink}$^\dagger$ and \uppercase{Martin Reuter}$^\ddagger$}

\address{Institute of Physics, University of Mainz\\
	Staudingerweg 7, D-55099 Mainz, Germany\\
	$^\dagger$E-Mail: nink@thep.physik.uni-mainz.de\\
	$^\ddagger$E-Mail: reuter@thep.physik.uni-mainz.de}

\begin{abstract}
We discuss the conceptual ideas underlying the Asymptotic Safety approach to the
nonperturbative renormalization of gravity. By now numerous functional renormalization
group studies predict the existence of a suitable nontrivial ultraviolet fixed point. We
use an analogy to elementary magnetic systems to uncover the physical mechanism behind
the emergence of this fixed point. It is seen to result from the dominance of certain
paramagnetic-type interactions over diamagnetic ones. Furthermore, the spacetimes of
Quantum Einstein Gravity behave like a polarizable medium with a ``paramagnetic''
response to external perturbations. Similarities with the vacuum state of Yang-Mills
theory are pointed out.
\end{abstract}


\vspace{0mm}
\bodymatter

\section{Introduction}
\label{sec:Intro}
The ultimate goal of the Asymptotic Safety program \cite{Weinberg} consists in
constructing a background independent quantum field theory of the gravitational
interaction and spacetime geometry. While its underlying degrees of freedom are carried
by the familiar tensor fields, such as the metric or tetrad and spin connection
variables, their quantization is intrinsically nonstandard, that is nonperturbative,
since the desired quantum theory will not be renormalizable within perturbation theory
presumably. For this reason, nonperturbative concepts and methods such as Wilson's modern
interpretation of ``renormalization'' play an important r\^{o}le. In this respect the
Asymptotic Safety program bears a certain resemblance to lattice field theory. In either
case an appropriate renormalization group fixed point is necessary in order to remove
the respective ultraviolet (UV) cutoff or, in lattice jargon, to take the continuum
limit. The idea behind Asymptotic Safety can be outlined as follows.

General relativity is considered an effective field theory approximation of an as yet
unknown full theory of quantum gravity. Such a full theory must be valid at all scales
and can be described by a scale dependent effective action. In principle this action can
comprise all possible field monomials respecting the required symmetries. Its scale
dependence is incorporated in running coupling constants, one for each field monomial.
Since this leads to an infinite number of couplings, one might fear a priori that all
of them need to be determined in an experiment so that the theory's predictivity is very
limited. However, the Asymptotic Safety scenario provides a possible way out. The basic
requirement is the existence of a fixed point under the renormalization group (RG)
operation which fixes the action in the UV. The inherent hypothesis, ultimately to be
proven, is that all those RG trajectories in the theory space of couplings which do
\emph{not} hit this fixed point in the UV are ``sick'' in the sense that they cannot
arise as the scale dependent effective action implied by a well-behaved complete
fundamental theory. In contrast to perturbation theory the fixed point may be nontrivial
here. One defines its ``UV-critical hypersurface'' as the set of all those points in the
infinite-dimensional theory space which are ``pulled'' into the fixed point by the
inverse RG flow: trajectories lying in this surface approach the fixed point for
increasing momentum scales. General arguments and known examples suggest that the
UV-critical hypersurface has a finite dimensionality. This dimensionality equals the
number of (infrared-) relevant couplings, i.e.\ couplings which get attracted to the
fixed point in the UV. The important point is that once the value of these (few)
couplings are known at some scale all other (irrelevant) couplings are fixed by requiring
an asymptotically safe theory, that is a trajectory which lies entirely in the
UV-critical hypersurface. By this means we achieve that, first, the couplings are
determined by a finite number of measurements rendering the theory predictive, and,
second, the UV behavior is unproblematic without any unphysical divergences.

In its form based upon the gravitational average action \cite{mr} the first step in the
Asymptotic Safety program consists in defining a coarse graining flow on an appropriate
theory space of action functionals. Then one searches for nontrivial fixed points of this
flow by means of functional renormalization group techniques. If such fixed points
exist, one must embark on the second step and try to construct a \emph{complete} RG
trajectory entirely within the theory space of well defined actions whereby the limit
corresponding to an infinite UV cutoff is taken at the fixed point in question. In the
successful case this trajectory defines a (candidate for a) nonperturbatively
renormalized quantum field theory whose properties and predictions can be explored then.
Furthermore, as the last step of the program one can use the RG trajectory in order to
construct a representation of the quantum theory in terms of a UV-regularized functional
integral; only then one will know the underlying Hamiltonian system which, implicitly,
got quantized by taking the UV limit at the fixed point computed \cite{elisa1}.

During the past decade a large number of detailed studies of the gravitational RG flow
has been performed and significant evidence for the viability of the Asymptotic Safety
program was found. In particular, all investigations carried out to date unanimously
agree on the existence of a non-Gaussian fixed point (NGFP) at which the infinite cutoff
limit can be taken \cite{mr,souma,frank1,oliver,NJP,livrev,reviews}.
However, as yet there is no general physical or mathematical understanding as to
\emph{why} this fixed point should exist, rendering Quantum Einstein Gravity (QEG)
nonperturbatively renormalizable. In fact, most of the existing investigations pick a
certain truncated theory space and then calculate the $\beta$-functions describing the RG
flow on it. Typically this step is technically extremely involved and so it would be very
desirable to gain a certain degree of intuitive understanding about the features of a
truncated action functional which are essential for exploring Asymptotic Safety and which
are not.

In this article we are going to describe the first step in this direction by identifying
a simple physical mechanism which, according to all average action-based studies of
Asymptotic Safety, seems to underlie the formation of the crucial non-Gaussian RG fixed
point in QEG. We shall demonstrate that it owes its existence to a \emph{predominantly
paramagnetic interaction of the metric fluctuations with an external gravitational
field.}

As we shall see, the key to understanding Asymptotic Safety is the following observation.
In a large class of well understood physical systems the pertinent quantum fluctuations
$\bp$ are governed by inverse propagators of the general form
\be
-D_\cA^2 + \mU(F_\cA)
\label{eqn:NonminOp}
\ee
where $D_\cA$ is the covariant derivative with respect to a certain connection $\cA$, and
$\mU$ denotes a matrix-valued potential depending on its curvature, $F_\cA$. The first
and the second term of the quadratic Lagrangian $\mathcal{L} = \frac{1}{2}\,\bp \big(
-D_\cA^2 + \mU(F_\cA) \big) \bp$ give rise to, respectively, diamagnetic-type and
paramagnetic-type interactions of the $\bp$'s with the background constituted by the
$\cA$ field. In the regime of the interest, the two types of interactions have an
antagonistic effect, but as the paramagnetic ones are much stronger than their
diamagnetic opponents they win and thus determine the qualitative properties of the
system. Well known examples of this ``paramagnetic dominance'' include the
susceptibility of magnetic systems, the screening of electric charges in QED, and the
antiscreening of color charges in Yang-Mills theory. Let us look at these systems in more
detail.

\noindent
\textbf{(i)} Nonrelativistic electrons in an external magnetic field are described by
the Pauli Hamiltonian
\be
H_\text{P} = \frac{1}{2m}\,(\mathbf{p}-e\mathbf{A})^2 + \mu_\text{B}\,
	\mathbf{B}\cdot \bm{\sigma}\,.
\label{eqn:PauliHam}
\ee
In the position space representation, $H_\text{P}$ becomes a nonminimal matrix
differential operator of the form (\ref{eqn:NonminOp}). The first term on the RHS of
(\ref{eqn:PauliHam}), essentially the gauge covariant Laplacian analogous to $D_\cA^2$ in
(\ref{eqn:NonminOp}), gives rise to the Landau diamagnetism of a free electron gas, while
the second term involving the ``curvature'' $\mathbf{B}$ of the ``connection''
$\mathbf{A}$ is the origin of the Pauli (spin) paramagnetism. The former is due to the
electrons' orbital motion, the latter to their spin alignment; they are characterized by
a negative ($\chi_\text{Landau-dia}<0$) and a positive ($\chi_\text{Pauli-para}>0$)
magnetic susceptibility, respectively. An important result is the relation between these
two quantities,
\be
\chi_\text{Landau-dia} = -\frac{1}{3}\, \chi_\text{Pauli-para}\,,
\label{eqn:DiaVsPara}
\ee
implying that it is always the \emph{paramagnetic} component which ``wins'' and
determines the overall sign of the total susceptibility:
$\chi_\text{mag}\equiv \chi_\text{Landau-dia}+\chi_\text{Pauli-para} >0$.

\noindent
\textbf{(ii)} The relativistic analog of (\ref{eqn:PauliHam}) is given by the square of
the Dirac operator $\sD\equiv \gamma^\mu D_\mu \equiv \gamma^\mu(\p_\mu-ieA_\mu)$, namely
\be
\sD^2 = D^2-\frac{i}{2}\,e\,\gamma^\mu \gamma^\nu F_{\mu\nu}\,.
\label{eqn:QEDoperator}
\ee
This differential operator, too, is nonminimal and has the structure
(\ref{eqn:NonminOp}). Again $D^2\equiv D^\mu D_\mu$ is responsible for the orbital
motion-related diamagnetic effects, and the non-de\-ri\-va\-tive term, which is
non-diagonal in spinor space, causes the pa\-ra\-mag\-ne\-tic effects.

\noindent
\textbf{(iii)} The same physics based distinction of orbital motion vs{.} spin alignment
effects can also be made for bosonic systems. Let us consider an {\sf SU}($N$) gauge
field $A_\mu^a \,$, $a=1,\cdots,N^2-1$, governed by the classical Yang-Mills Lagrangian
$\propto (F_{\mu\nu}^a)^2$. If we expand $A_\mu^a$ at some background field
$\bar{A}_\mu^a$, small fluctuations about this background, $\delta A_\mu^a$, are described
by a quadratic action of the form $\int \delta A_\mu^a(\cdots)\delta A_\nu^b$. In Feynman
gauge the kernel $(\cdots)$ is given by the nonminimal operator, again of the type
(\ref{eqn:NonminOp}),
\be
\left(\bar{D}^2\right)^{ab}\delta_\mu^{~\nu} - 2ig\bar{F}^{ab~\nu}_{~~\mu}\,.
\label{eqn:YMoperator}
\ee
Here $\bar{D}^2\equiv \bar{D}^\mu \bar{D}_\mu$ and $\bar{F}^{ab~\nu}_{~~\mu}$ are built
from the background field. The fluctuations have two qualitatively different interactions
with the background field: an orbital one $\propto \int\delta A \, \bar{D}^2 \delta A$
related to their spacetime dependence, and an ultralocal one $\propto\int \delta A \,
\bar{F} \delta A$ which is sensitive to the orientation of the fluctuations relative to
the background in color space. In Yang-Mills theory, too, there are specific effects
which can be attributed to the first, ``diamagnetic'', and the second, ``paramagnetic'',
term separately \cite{Nielsen}.

A well known example \cite{huang-book,dittrich} is asymptotic
freedom \cite{PolitzerGrossWilczek}. In fact, the one-loop Yang-Mills $\beta$-function
can be decomposed as $\beta_{g^2} = - \frac{N}{24\pi^2}\,\big[\,12-2+1 \,\big]\,g^4$,
where the ``+12'' is due to the fluctuations' paramagnetic interaction, the ``-2'' stems
from the diamagnetic one, and the ``+1'' comes from the Faddeev-Popov ghosts. The para-
and diamagnetic contributions come with an opposite sign, but since the former are six
times bigger than the latter, it is the paramagnetic interaction that determines the
overall negative sign of $\beta_{g^2}$. In this respect $\beta_{g^2}$ is analogous to the
magnetic susceptibility $\chi_\text{mag}$ whose sign is also determined by the
competition of para- and diamagnetic effects, the clear winner being paramagnetism.
Thus we can say that in Yang-Mills theory \emph{asymptotic freedom is due to the
predominantly paramagnetic interaction of gauge field fluctuations with external fields.}

\noindent
\textbf{(iv)} As we shall see, the Asymptotic Safety of QEG is, in the sense of
this magnetic analogy, very similar to the asymptotic freedom of Yang-Mills theory, the
main difference being that the Gaussian fixed point implicit in perturbative
renormalization is replaced by a nontrivial one now. The similarities are most clearly
seen in the Einstein-Hilbert truncation of the QEG theory space \cite{mr}. The dynamics
of fluctuations $h_\mn(x)$ about a prescribed metric background $\bar{g}_\mn(x)$ is given
by a quadratic action $\propto \int\sgb\; h_\mn(\cdots)\,h^{\rho\sigma}$ whose kernel
$(\cdots)$ is found by expanding the Einstein-Hilbert action to second order. The result
with a harmonic gauge fixing is again a nonminimal matrix differential operator with a
clear separation of ``dia-'' vs. ``paramagnetic'' couplings to the background:
\be
-\bar{K}^\mn_{~~\rho\sigma} \, \bar{D}^2 + \bar{U}^\mn_{~~\rs}\,.
\label{eqn:GravOperator}
\ee
Here $\bar{D}^2\equiv\gb^\mn \bar{D}_\mu\bar{D}_\nu$, where $\bar{D}_\mu$ is the
covariant derivative with respect to the Levi-Civita connection given by $\gb_\mn$,
and $\bar{U}^\mn_{~~\rs}$ is a tensor built from the background's curvature tensor,
\be
\begin{split}
\bar{U}^\mn_{~~\rs} = \frac{1}{4}\left[\delta_\rho^\mu \delta_\sigma^\nu +
	\delta_\sigma^\mu \delta_\rho^\nu - \gb^\mn \gb_\rs \right] \left(\bar{R}-2\Lambda_k\right)
	+ \frac{1}{2}\left[\gb^\mn \bar{R}_\rs + \gb_\rs \bar{R}^\mn\right]& \\
	- \frac{1}{4}\left[\delta_\rho^\mu\bar{R}^\nu_{~\sigma} + \delta_\sigma^\mu\bar{R}^\nu_{~\rho}
	+ \delta_\rho^\nu\bar{R}^\mu_{~\sigma} + \delta_\sigma^\nu\bar{R}^\mu_{~\rho}\right]
	- \frac{1}{2}\left[\bar{R}^{\nu~\mu}_{~\rho~\sigma} + \bar{R}^{\nu~\mu}_{~\sigma~\rho}\right]&.
\end{split}
\label{eqn:UGrav}
\ee
Furthermore, $\bar{K}^\mn_{~~\rs} = \frac{1}{4}\left[\delta_\rho^\mu \delta_\sigma^\nu
+\delta_\sigma^\mu \delta_\rho^\nu - \gb^\mn \gb_\rs \right]$.
The r\^{o}le of $\beta_{g^2}$ in 4D Yang-Mills theory is played by the anomalous
dimension $\eta_N$ of Newton's constant now. Here, too, we will see that it is possible
to disentangle dia- and paramagnetic contributions. Again they come with opposite signs,
and the paramagnetic effects turn out much stronger than their diamagnetic competitors or
the ghosts.

As a consequence, the negative sign of $\eta_N$ governing the RG running of Newton's
constant, crucial for Asymptotic Safety and gravitational antiscreening, originates from
the paramagnetic interaction of the metric fluctuations with their background (or
``condensate''). The diamagnetic effects counteract the antiscreening trend and the
formation of an NGFP, but they are too weak to overwhelm the paramagnetic ones. This is
what we shall call \emph{paramagnetic dominance.}
\medskip

Before presenting the details of this mechanism we briefly summarize the essentials of
the calculational method we employ. The scale dependence due to the renormalization or
``coarse graining'' processes is studied by means of the effective average action
$\Gamma_k$ \cite{avact}. The basic feature of this action functional consists in
integrating out all quantum fluctuations of the underlying fields from the UV down to an
infrared cutoff scale $k$. In a sense, $\Gamma_k$ can be considered the generating
functional of the 1PI correlation functions that take into account the fluctuations of
all scales larger than $k$. Consequently, for $k=0$ it coincides with the usual effective
action, $\Gamma_{k=0}=\Gamma$. On the other hand, in the limit $k\rightarrow\infty$,
$\Gamma_k$ approaches the bare action $S$, apart from a simple, explicitly known
correction term \cite{elisa1}.

Starting from the functional integral definition of $\Gamma_k$ one may investigate its
scale dependence by taking a $k$-derivative, which results in the exact functional
renormalization group equation (FRGE) \cite{avact,YM-EAA}
\be
\p_t \Gamma_k = \frac{1}{2}\, \text{STr}\left[ \left( \Gamma_k^{(2)}+ \Rk \right)^{-1}
	\p_t \Rk \right] \,.
\label{eqn:FRGE}
\ee
Here we introduced the RG time $t=\ln k$. The differential operator $\Rk$ in
(\ref{eqn:FRGE}) comprises the infrared cutoff: in the corresponding path integral the
bare action $S$ is replaced by $S+\Delta_k S$ where the cutoff action added is quadratic
in the fluctuations $\phi$, $\Delta_k S \propto \int \phi\Rk\phi$. Typically $\Rk$ has
the structure  $\Rk \propto k^2\RN(-\Delta)$ where $\Delta$ is an appropriate
Laplace-type operator and $\RN$ a scalar cutoff shape function interpolating between
$\RN(0)=1$ and $\RN(\infty)=0$. Furthermore,
$\Gamma_k^{(2)}$ denotes the ``matrix'' of second functional derivatives with respect to
the dynamical fields. The functional supertrace in (\ref{eqn:FRGE}) includes a trace over
all internal indices, and a sum over all dynamical fields, with an extra minus sign for
the Grassmann-odd ones.

The effective average action ``lives'' in the infinite dimensional ``theory space'' of
all action functionals depending on a given set of fields and respecting the required
symmetries. Its RG evolution determined through eq.\ (\ref{eqn:FRGE}) amounts to a curve
$k \mapsto \Gamma_k$ in theory space. Since the FRGE leads in general to a system of
infinitely many coupled differential equations, one usually has to resort to truncations
of the theory space. For this purpose $\Gamma_k$ is expanded in a basis of field
monomials $P_\alpha[\,\cdot\,]$, i.e.
$\Gamma_k[\,\cdot\,]=\sum_\alpha c_\alpha(k) P_\alpha[\,\cdot\,]$,
but then one truncates the sum after a finite number of terms. Thus the scale dependence
of $\Gamma_k$ is described by finitely many running couplings $c_\alpha(k)$. Projecting
the RHS of (\ref{eqn:FRGE}) onto the chosen subspace of theory space the resulting system
of differential equations for the couplings remains finite, too. In this way we obtain
the $\beta$-functions for the couplings $c_\alpha$.

In order to illustrate the r\^{o}le of ``paramagnetic dominance'' we shall always use the
simplest truncation possible for the respective system, for instance the Yang-Mills
action $\propto \int(F_\mn^a)^2$ for an {\sf SU}$(N)$ gauge theory, or the
Einstein-Hilbert action in the case of gravity, each one furnished with running
couplings.

\medskip
The remaining sections in this article are organized as follows. Section \ref{sec:QED-YM}
demonstrates the idea of distinguishing between the two ``magnetic'' contributions with
the help of two examples: fermions in QED and gauge bosons in Yang-Mills theory. In
section \ref{sec:QEG} we perform the analogous analysis for gravity and focus on the
question which terms render QEG, in the Einstein-Hilbert truncation, asymptotically safe.
Finally, section \ref{sec:QEGvacuum} proposes the interpretation of the QEG spacetimes as
a polarizable medium, and section \ref{sec:conclusion} contains the conclusions.

\section{Paramagnetic dominance and paramagnetic vacua}
\label{sec:QED-YM}
This section is meant to demonstrate our FRGE based ``dia vs.\ para'' decomposition by
means of two well known examples: QED and Yang-Mills theory.
 
By calculating the interaction energy between two (generalized) charges it is possible
to define analogs of the electric and magnetic susceptibility also for other field
theories than electrodynamics, for instance Yang-Mills theory.
From a renormalization point of view this can be used to establish a connection between
the susceptibility and the $\beta$-function. Let us consider a massless charged field
with spin $S$ and renormalized charge $g$. The lowest order of the $\beta$-function for
$g^2$ is quartic in $g$, so that one can expand $\beta_{g^2}=\beta_0 g^4+\ord(g^6)$.
Then one finds a relation for the magnetic susceptibility,
$\chi_\text{mag}\propto \beta_0$, where $\beta_0$ is given by \cite{Nielsen}
\be
\beta_0 = - \frac{(-1)^{2S}}{4\pi^2} \,\left[ \Big\{ (2S)^2 \Big\}_\text{para} +
	\Big\{ - \textstyle\frac{1}{3} \Big\}_\text{dia} \, \right] \,.
\label{eqn:NielsenHughes}
\ee
Here the first term, $(2S)^2$, is due to the ``paramagnetic'' interaction, while the
$-\frac{1}{3}$ is the ``diamagnetic'' contribution. The notation involving the curly
brackets to separate the different magnetic effects will be employed in the following
sections, too. One can check that, for spin-$\frac{1}{2}$ fermions, eq.\
(\ref{eqn:NielsenHughes}) reduces to $\beta_0^\text{QED} = \frac{1}{4\pi^2}\left[
\{1\}_\text{para} + \{-\frac{1}{3}\}_\text{dia} \right]$, reproducing relation
(\ref{eqn:DiaVsPara}): $\beta_0^\text{dia}=-\frac{1}{3}\beta_0^\text{para}$. Likewise,
for QCD without fermions, eq.\ (\ref{eqn:NielsenHughes}) yields
$\beta_0^\text{QCD}=-\frac{1}{8\pi^2}\big[\{12\}_\text{para}+\{-1\}_\text{dia}\big]$.

\subsection{Paramagnetic dominance in QED}
\label{sec:QED}
In Ref.~\refcite{andreas1} the QED $\beta$-function was derived up to the lowest order in
the coupling constant $e$ by means of functional RG techniques. The separation into the
different magnetic contributions was done when encountering the operator
$-\slashed{D}^2=-D^2+\frac{i}{2}\,e\gamma^\mu\gamma^\nu F_\mn$: the first term induces
diamagnetic effects, while the second one is responsible for paramagnetic effects. This
leads to the $\beta$-function
\be
\p_t e^2 = \beta_{e^2} = \frac{1}{4\pi^2}\, \Bigg[ \bigg\{1\bigg\}_\text{para} +
	\bigg\{-\frac{1}{3}\bigg\}_\text{dia} \;\Bigg]\, e^4 \,,
\label{eqn:betaQED}
\ee
in perfect agreement with the general formula (\ref{eqn:NielsenHughes}) for $S=
\frac{1}{2}$. We clearly see the relation $\beta_{e^2}^\text{dia} = -\frac{1}{3}\,
\beta_{e^2}^\text{para}$. The positive sign of $\beta_{e^2}$ is a crucial feature of QED.
It is due to its screening character.
Since it is the paramagnetic term in (\ref{eqn:betaQED}) that dictates the overall sign,
we can conclude that the qualitative properties of QED, particularly its asymptotic
behavior, are determined by paramagnetism. We emphasize that these findings, based on
(\ref{eqn:betaQED}), are \emph{universal}, i.e.\ they are independent of the cutoff shape
function we choose.

\subsection{Paramagnetic dominance in Yang-Mills theory}
\label{sec:YM}
Now we transfer the concepts employed above for QED to the non-Abelian case, and
investigate in particular the origin of asymptotic freedom in Yang-Mills theory. We keep
the spacetime dimension $d$ arbitrary. For $d\neq 4$ the Yang-Mills coupling is
dimensionful and so the $\beta$-function of its dimensionless counterpart contains the
classical scaling dimension $d-4$ besides the anomalous dimension $\eta_F$. Thus we have
$\p_t g^2 = \beta_{g^2} \equiv (d-4+\eta_F)g^2$.
Since it is $\eta_F$ that comprises the quantum effects we are interested in, we shall
discuss the different ``magnetic'' contributions at the level of $\eta_F$ rather than the
$\beta$-function.

Within the nonperturbative setting the calculation proceeds as follows. If we employ the
background formalism \cite{dewitt-books}, i.e.\ split the dynamical gauge field $A_\mu^a$
into a rigid background $\bar{A}_\mu^a$ and a fluctuation $a_\mu \equiv \delta A_\mu^a$,
the propagation of the fluctuating field is crucially influenced by its interaction with
the background. In this regard the background assumes the r\^{o}le of an external color
magnetic field that couples to the fluctuations and probes their properties. Our goal is
to determine the scale dependence of the corresponding coupling constant.

Choosing the gauge group {\sf SU}$(N)$ we construct gauge invariant combinations of the
gauge field $A_\mu^a$ as candidates for appropriate action monomials. Here it is
sufficient to follow Ref.\ \refcite{YM-EAA} and consider a simple truncation for $\Gamma_k$
which consists of the usual Yang-Mills action, equipped with a scale dependent prefactor,
plus a gauge fixing term:
\be
\Gamma_k[A,\bar{A}\,] = \int \ddx \left\{ \frac{1}{4}\, \ZFk\, F_\mn^a[A] F_a^\mn[A]
	+ \frac{\ZFk}{2\alpha_k} \left[ D_\mu[\bar{A}](A^\mu-\bar{A}^\mu) \right]^2 \right\} \,.
\label{eqn:GammaYM}
\ee
The field strength tensor is given by
$F_\mn^a[A] = \p_\mu A_\nu^a - \p_\nu A_\mu^a + \bar{g}\,f_{bc}{}^a\, A_\mu^b A_\nu^c$
with the bare charge $\bar{g}$ and the structure constants $f_{bc}{}^a$. The gauge fixing
parameter $\alpha_k$ will be approximated by the $k$-independent value $\alpha=1$ in the
following.

For truncations of the type (\ref{eqn:GammaYM}) the general, exact FRGE for Yang-Mills
fields \cite{YM-EAA} boils down to the following decomposed form which treats gauge boson
and ghost contributions separately:
\be
\begin{split}
\p_t\Gamma_k[A,\bar{A}\,] = &\;\frac{1}{2}\, \Tr\left\{ \left( \Gamma_k^{(2)}[A,\bar{A}\,]
	+ \Rk[\bar{A}\,] \right)^{-1} \p_t \Rk[\bar{A}\,] \right\} \\
& - \Tr \left\{ \left( -\bar{D}^2	+ \Rk^\text{gh}[\bar{A}\,] \right)^{-1}
	\p_t \Rk^\text{gh}[\bar{A}\,] \right\} \,.
\end{split}
\label{eqn:FRGE-YM}
\ee
After taking the second functional derivative in (\ref{eqn:FRGE-YM}) we identify
$\bar{A}=A$, project the traces onto the functional $\int\ddx F_\mn^a[A] F_a^\mn[A]$, and
deduce the running of $\ZFk$. With the gauge fields identified, $\Gamma_k^{(2)}$ reduces
to
\be
\Gamma_k^{(2)}[A] \equiv \frac{\delta^2}{\delta A^2} \Gamma_k[A,\bar{A}\,]
	\Big|_{\bar{A}=A} = \ZFk \left( -D^2 + 2i\bar{g}\,F \,\right)
	\,.
\label{eqn:YMGammatwo}
\ee
We observe that the operator (\ref{eqn:YMGammatwo}) has a similar form as its QED analog,
$-\slashed{D}{}^2=-D^2+\frac{i}{2}\,e\,\gamma^\mu\gamma^\nu F_\mn$. Thus an obvious
notion of ``dia-'' vs.\ ``paramagnetic'' contributions suggests itself: the first term of
the RHS in (\ref{eqn:YMGammatwo})
represents diamagnetic interactions, and the second, nonminimal, term paramagnetic
ones. The only difference compared to the fermions in QED occurs due to the additional
ghost term in the FRGE. Since the ghost analog of (\ref{eqn:YMGammatwo}) is only the
minimal operator $-\bar{D}^2$, its induced effects will be referred to as
``ghost-diamagnetic'' in the following.

We expand the traces in (\ref{eqn:FRGE-YM}) as a heat kernel series \cite{heat-kernel},
equate the coefficients of $\int\ddx F_\mn^a F_a^\mn$, and obtain a differential
equation describing the scale dependence of $\ZFk\,$.\footnote{See Ref.~\refcite{YM-EAA}
for the details of the calculation.} In terms of the dimensionless renormalized charge
$g^2 \equiv k^{d-4}\ZFk^{-1}\, \bar{g}^2$ this yields the following $\ord(g^2)$ result 
for the anomalous dimension $\eta_F \equiv -\p_t \ln \ZFk \,$:
\be
\eta_F(g) = - \frac{1}{3}\,(4\pi)^{-d/2}\, N \, \Phi_{d/2-2}^1(0) \Big[ \,
	\big\{24\big\}_\text{para} + \big\{-d\big\}_\text{dia}	+ \big\{2\big\}_\text{ghost-dia}
	\Big]	g^2 \,.
\label{eqn:etaYM}
\ee
The factor $\Phi_{d/2-2}^1(0)$ in (\ref{eqn:etaYM}) denotes one of the standard threshold
functions \cite{mr}, evaluated at vanishing argument. The threshold functions are all
\emph{positive}, in particular, concerning eq.\ (\ref{eqn:etaYM}), $\Phi_{d/2-2}^1(0)>0$
for any $d$. In a generic dimension the numerical value of $\Phi_{d/2-2}^1(0)$ depends on
the shape function $\RN$. The case $d=4$ is special because of its universality:
$\Phi^1_0(0)=1$ for any $\RN$.

We emphasize the importance of (\ref{eqn:etaYM}). For all $d<24$
we find the paramagnetic part to be dominant. With regard to relative signs, the
diamagnetic effect counteracts the paramagnetic and the ghost one. However, the
diamagnetic contribution is subdominant up to the critical dimension $d=26$ which has
$\eta_F = 0$. Hence, for $d < 26$ the anomalous dimension is negative, and this is
basically due to the paramagnetic term. In turn, it is this sign that determines the
qualitative behavior of the coupling $g$ at high energies. Therefore, one can say that
\emph{paramagnetism decides about whether or not the Yang-Mills theory is asymptotically
free/safe}.

Finally, we focus on the case $d=4$. Then $\eta_F$ becomes universal since $\Phi_0^1(0)
= 1$ for any cutoff, and so we obtain $\eta_F = - \frac{N}{24\pi^2}\,\big[\,
\{12\}_\text{para} + \{-2\}_\text{dia} + \{1\}_\text{ghost-dia}\, \big]g^2$.
The crucial overall minus sign driving $g$ to zero in the high energy limit results from
the first term of the sum. Thus we can conclude for four-dimensional Yang-Mills theory
that \emph{asymptotic freedom occurs only due to the paramagnetic interactions}.
\medskip

Let us recapitulate. We considered inverse propagators of the form $-\bar{D}^2+U$ with a
potential term $U$. They consist of two parts, a minimal one of Laplace type and a
nonminimal one. The effects induced by these different parts, in particular their impact
on the $\beta$-function, are called dia- and paramagnetic, respectively. We found the
latter to prevail in QED and Yang-Mills theory. Dia- and paramagnetism, in this sense,
correspond to rather different types of interactions the quantized field fluctuations
have with their classical background: via their spacetime modulation, measured by
$\bar{D}^2$ in the ``dia'' case, and by aligning their internal degrees of freedom to the
external field in the ``para'' case.

\subsection{The vacuum as a magnetic medium}
\label{sec:vacuum}
The results of the previous subsections are a confirmation of eq.\
(\ref{eqn:NielsenHughes}) for the cases with spin $S=\frac{1}{2}$ and $S=1$,
respectively. More generally, this formula is valid for any massless field of spin $S$
which carries nonzero Abelian or non-Abelian charge and has a $g$-factor of exactly $2$.
As for paramagnetic dominance we note an important point about relation
(\ref{eqn:NielsenHughes}): Even though for any spin $S \geq\frac{1}{2}$
the paramagnetic contribution to $\beta_0$ exceeds the diamagnetic one (which possesses
the opposite sign), the total sign of $\beta_0$ alternates with $S$ due to the overall
factor $(-1)^{2S}$, being a consequence of the Feynman rule that fermion loops come with
an extra minus sign. Without this extra factor, QED, too, would be asymptotically free as
both in QED and in Yang-Mills theory the ``para'' contribution to the $\beta$-function
overrides the ``dia'' one.

Asymptotic freedom can be understood by regarding the vacuum state of the quantum field
theory as a magnetic medium and analyzing its response to an external magnetic field
\cite{Nielsen,huang-book}. Defining the (color) electric and
magnetic permeabilities, $\varepsilon$ and $\mu$, respectively,\cite{andreas1} one finds
$\varepsilon\mu=1$ as a consequence of Lorentz invariance. If $\mu<1$ a medium is
referred to as diamagnetic, and if $\mu>1$ it is called paramagnetic. It is convenient to
base the discussion on the corresponding susceptibilities $\chi_\text{el}$ and
$\chi_\text{mag}$, given by $\varepsilon\equiv 1+\chi_\text{el}$ and
$\mu\equiv 1+\chi_\text{mag}$. For $\varepsilon$ and $\mu$ close to unity Lorentz
invariance hence leads to the relation $\chi_\text{el}\approx - \chi_\text{mag}$.

A homogeneous, isotropic medium, e.g.\ the vacuum of a quantum field theory, is
\emph{screening} (color) electric charges if $\varepsilon > 1$, $\chi_\text{el} > 0$.
This implies $\mu < 1$, $\chi_\text{mag} < 0$ in relativistic case, and so the vacuum
represents a \emph{diamagnetic medium}.

If we have $\varepsilon < 1$, $\chi_\text{el} < 0$ instead, the medium is
\emph{antiscreening} electric charges, and Lorentz invariance implies $\mu > 1$,
$\chi_\text{mag} > 0$. The vacuum state is a \emph{paramagnetic medium} then.

Here we employ the usual terminology of referring to a medium as dia- \mbox{(para-)}
magnetic when $\mu < 1$ ($\mu > 1$). We emphasize that a priori this notion has nothing
to do with the dia- vs.\ paramagnetic \emph{interactions} discussed previously. Only in
the nonrelativistic theory for standard magnetic materials ``paramagnetic dominance''
entails a ``paramagnetic medium''. However, in the present generalized context \emph{it
is possible that a quantum field theory has a vacuum state which behaves as a diamagnetic
medium even though the paramagnetic interactions dominate}.

This possibility is directly related our discussion about the factor $(-1)^{2S}$
in $\beta_0$, given by the formula (\ref{eqn:NielsenHughes}). In fact, if one determines
the effective Lagrangian for the theory this formula applies to, and computes the
corresponding field dependent magnetic susceptibility from it, one obtains
\cite{Nielsen,huang-book}
\be
\chi_\text{mag}(B) = - \frac{1}{2}\,\beta_0\, g^2\, \ln
	\left( \frac{\Lambda^2}{gB} \right) \,,
\label{eqn:chiYM}
\ee
where $\Lambda$ is a UV cutoff, and the normalization is chosen such that
$\chi_\text{mag}(B=\Lambda^2/g) = 0$. Lowering $B$ below $\Lambda^2/g$ means integrating
out the modes with eigenvalues lying in the interval $[gB,\Lambda^2]$. This renders
$\chi_\text{mag}$ nonzero, and according to eq.\ (\ref{eqn:chiYM}) we have:
$\beta_0 < 0 \; \Leftrightarrow\; \chi_\text{mag} > 0 \;$(paramagnetic medium), and,
equivalently
$\beta_0 > 0 \; \Leftrightarrow\; \chi_\text{mag} < 0 \;$(diamagnetic medium).

As a consequence, by virtue of $\beta_0^\text{QED} > 0$ the QED vacuum is a diamagnetic
medium, while, since $\beta_0^\text{YM} < 0$, the vacuum state of the non-Abelian gauge
bosons in Yang-Mills theory is a paramagnetic one. But concerning the relative strength
of the two interactions we have ``paramagnetic dominance'' in both cases!

\section{Asymptotic Safety and paramagnetic dominance in QEG}
\label{sec:QEG}
In this section we transfer the concept of separating magnetic contributions introduced
above to Quantum Einstein Gravity. We reveal a general mechanism behind its
asymptotically safe behavior which is at the heart of our approach to a quantized theory
of gravity.

Asymptotic Safety has been confirmed for many different models and truncations
\cite{NJP,livrev,reviews} after the first introduction of the gravitational average
action \cite{mr}. The system of RG equations for the running couplings possesses a
non-Gaussian fixed point (NGFP) for any truncation considered. However, the question
about the physical explanation for the emergence of an NGFP has been raised only
recently \cite{andreas1}. Here we summarize the main ideas outlined in Ref.\
\refcite{andreas1}, proposing generalized dia- and paramagnetic interactions of the
quantum fluctuations with their background.

Using the background field formalism the dynamical metric, $\gamma_\mn$, is split into a
sum of a fixed but arbitrary background metric $\bar{g}_\mn$ and the fluctuation
$h_\mn$; its expectation value reads $g_\mn \equiv \langle\gamma_\mn \rangle =
\bar{g}_\mn + \bar{h}_\mn$ with $\bar{h}_\mn \equiv \langle h_\mn\rangle$. The average
action $\Gamma_k[\bar{h}_\mn;\bar{g}_\mn]$, or equivalently $\Gamma_k[g_\mn,\bar{g}_\mn]
\equiv \Gamma_k[g_\mn-\bar{g}_\mn;\bar{g}_\mn]$, thus depends on two independent
arguments. After expanding $\Gamma_k$ in terms of $\bar{h}_\mn$ we encounter interaction
terms of the metric fluctuations with the background field at any order in $\bar{h}_\mn$.
From the point of view of these fluctuations (``gravitons'') one can regard the
background geometry as an external ``magnetic'' field which polarizes the quantum vacuum
of the ``$h_\mn$-particles'', and gives rise to a corresponding susceptibility. This
analogy to Yang-Mills theory suggests a separation of the dia- and paramagnetic
mechanisms also here by disentangling kinetic (orbital) and alignment effects.

\subsection{Dia- vs.\ paramagnetism in the Einstein-Hilbert truncation}
\label{sec:EHDiaPara}
Studying QEG within the Einstein-Hilbert truncation we derive the $\beta$-functions
along the lines of Ref.~\refcite{mr}, but after having computed the inverse propagator
$\Gamma_k^{(2)}$ the separation into the magnetic components is performed.

Our truncation ansatz comprises the Einstein-Hilbert action, a gauge fixing term
and a ghost action, $\Gamma_k = \Gamma_k^\text{EH} + \Gamma_k^\text{gf} +
\Gamma_k^\text{gh} \equiv \breve{\Gamma}_k + \Gamma_k^\text{gh}$, where the scale
dependence is described by a running Newton constant $G_k$ and a cosmological constant
$\Lambda_k$:
\be
\breve{\Gamma}_k[g,\bar{g}] =\, \frac{1}{16\pi G_k} \int\ddx\, \sg \left( -R[g] 
		+ 2\Lambda_k \right) + \Gamma_k^\text{gf} \,.
\label{eqn:Einstein-Hilbert}
\ee
Here $\Gamma_k^\text{gh}$ coincides with the classical, scale independent ghost action.
Employing the harmonic coordinate condition to fix the gauge the resulting Faddeev-Popov
operator $\mathcal{M}$ reads
$\mathcal{M}[g,\bar{g}]^\mu_{~\nu} = \gb^{\mu\rho} \gb^{\sigma\lambda} \bar{D}_\lambda
	(g_{\rho\nu}D_\sigma + g_{\sigma\nu}D_\rho) - \gb^{\rho\sigma} \gb^{\mu\lambda}
	\bar{D}_\lambda g_{\sigma\nu}D_\rho \,$.
	
The corresponding FRGE assumes a decomposed form, with one trace for the gravitons and
another for the ghosts:
\be
\begin{split}
\p_t\Gamma_k[g,\bar{g}] = \frac{1}{2}\, &\Tr\left[\left(\breve{\Gamma}_k^{(2)}[g,\bar{g}]
	+ \Rk^\text{grav}[\bar{g}] \right)^{-1} \p_t\Rk^\text{grav}[\bar{g}] \right]\\
	- &\Tr \left[ \left( -\mathcal{M}[g,\bar{g}] + \Rk^\text{gh}[\bar{g}]\right)^{-1}
	\p_t\Rk^\text{gh}[\bar{g}]\right] \,.
\label{eqn:FRGE-QEG}
\end{split}
\ee
After having determined $\breve{\Gamma}_k^{(2)}$ we set $\gb = g$. In this way we obtain
the nonminimal operator
\be
\left(\breve{\Gamma}_k^{(2)}[g,\gb]\right)^\mn_{~~\rs}\Big|_{g=\gb} = \frac{1}{32\pi G_k}
	\,\Big( -\bar{K}^\mn_{~~\rs} \bar{D}^2 + \bar{U}^\mn_{~~\rs} \Big) \,,
\label{eqn:Gamma2Grav}
\ee
with $\bar{K}^\mn_{~~\rs}$ and $\bar{U}^\mn_{~~\rs}$ as given in eqs.\
(\ref{eqn:GravOperator}), (\ref{eqn:UGrav}). The Faddeev-Popov operator, too, becomes a
nonminimal operator that involves the covariant Laplacian $-\bar{D}^2$ and an additional
potential term: $-\mathcal{M}[g,\bar{g}]^\mu_{~\nu}|_{g=\gb}=\delta^\mu_\nu\big(-\bar{D}^2
	-d^{-1}\bar{R}\big)$.

By analogy with QED and Yang-Mills theory we can identify the different ``magnetic''
components in these two operators now. Considering the one in (\ref{eqn:Gamma2Grav}) its
first term gives rise to \emph{diamagnetic} interactions, while the second one induces
\emph{paramagnetic} effects. Similarly, contributions from the first part in the
Faddeev-Popov operator are referred to as \emph{ghost-diamagnetic}, those coming from
the second term as \emph{ghost-paramagnetic}.

Note that the FRGE (\ref{eqn:FRGE-QEG}) is valid for any background metric $\bar{g}_\mn$.
Therefore, it is reasonable to choose the background such that the computation of the
running of the couplings gets simplified. We assume a maximally symmetric space here
since it still allows for an identification of the terms used in our truncation
\cite{mr}. Then all curvature terms in $\bar{U}^\mn_{~~\rs}$ are proportional to $R$. As
a consequence the denominators of both traces on the RHS of eq.\ (\ref{eqn:FRGE-QEG}) are
of the general form $(\mathcal{A}+CR)^{-1}$, where $C$ is a constant and $\mathcal{A}$
denotes a function of the Laplacian $-D^2$ that does not contain any curvature terms.
Expanding
\be
(\mathcal{A}+CR)^{-1} = \mathcal{A}^{-1} - C\mathcal{A}^{-2}\,R+\ord(R^2) \,,
\ee
we can identify $\mathcal{A}^{-1}$ as the diamagnetic part, while the term proportional
to $R$ is the paramagnetic contribution.

The evaluation of the traces is then done as in Ref.~\refcite{mr}. One finally obtains
two differential equations describing the running of $\Lambda_k$ and $G_k$, or of the
corresponding dimensionless coupling constants, $\lambda_k \equiv k^{-2} \Lambda_k$ and
$g_k \equiv k^{d-2}G_k$, respectively.

The flow equation for the cosmological constant reads
\ba
\p_t \lambda_k = \beta_\lambda(g_k,\lambda_k) \equiv\, &\big[\eta_N(g_k,\lambda_k)-2\big]
	\lambda_k + 2\pi g_k (4\pi)^{-d/2} \Big[ 2d(d + 1)\, \Phi_{d/2}^1(-2\lambda_k)
	\nonumber\\
	&-d(d + 1)\, \eta_N(g_k,\lambda_k)\, \widetilde\Phi_{d/2}^1(-2\lambda_k)
	- 8d\, \Phi_{d/2}^1(0)\Big] \,.
\label{eqn:betalambda}
\end{align}
The threshold functions $\Phi$, $\widetilde\Phi$ in (\ref{eqn:betalambda})
are the same as in Ref.~\refcite{mr}. In order to identify dia- and paramagnetic
contributions to the anomalous dimension $\eta_N\equiv \p_t\ln G_k$ we employ the
separation rule outlined above. The result can be written as
\be
\eta_N(g,\lambda) = \frac{g \, B_1(\lambda)}{1-g \, B_2(\lambda)} \,.
\label{eqn:etaN}
\ee
As we will show below, the term $gB_2(\lambda)$ in the denominator of (\ref{eqn:etaN}) is
irrelevant for our qualitative discussion. In contrast, the function $B_1$ in the
numerator is of fundamental importance. It contains ``dia'' and ``para'' terms from
both gravitons and ghosts:
$B_1(\lambda) \equiv\, \frac{1}{3}\, (4\pi)^{1-\frac{d}{2}} \big[\, \big\{
	d(d+1) \Phi_{d/2-1}^1(-2\lambda) \big\}_\text{dia}
	+ \big\{ -4d\, \Phi_{d/2-1}^1(0) \big\}_\text{ghost-dia}\linebreak
	\qquad + \big\{ -6d(d-1)\, \Phi_{d/2}^2(-2\lambda) \big\}_\text{para}
	+ \big\{-24\, \Phi_{d/2}^2(0) \big\}_\text{ghost-para} \;\big]$.
The RG equation of Newton's constant involves this anomalous dimension:
\be
\p_t g_k = \beta_g(g_k,\lambda_k) \equiv \big[ d-2+\eta_N(g_k,\lambda_k) \big]\, g_k \,.
\label{eqn:betag}
\ee

As already mentioned above, $\eta_N$ consists of two parts of different significance.
The numerator $g\, B_1(\lambda)$ decides on the qualitative behavior of $\eta_N$. In
particular it determines the overall sign. By contrast, the denominator
$1-g\, B_2(\lambda)$ assumes the r\^{o}le of a correction term only. 
Due to the singularity it gives rise to at $g=1/B_2(\lambda)$ it delimits the theory
space in the $g$-$\lambda$ plane \cite{frank1}. However, away from this boundary
singularity it does not change significantly the leading order behavior of $\eta_N$ given
by the numerator.

The important features for our analysis are contained in $g\, B_1(\lambda)$ alone since
we focus on the sign of $\eta_N$. We may thus perform the expansion $\eta_N= g\,
B_1(\lambda) + \ord(g^2)$ and retain the term linear in $g$ only.\footnote{Despite this
expansion in $g$, we are still aiming at a \emph{nonperturbative} renormalization of QEG,
meaning that the continuum limit is taken at an NGFP with $g_* \neq 0$ rather than the
trivial fixed point of perturbation theory, $g_* = 0$. If the NGFP, in some scheme, has
a small but nonzero $g_* B_2(\lambda_*)$ it might well be possible to find it in a small
coupling expansion. However, the latter may not be confused with ``perturbation theory''
in the sense of ``perturbative renormalization''.} The various magnetic contributions to
$\eta_N$ then follow from the separation we found for $B_1(\lambda)$:
\ba
&\eta_N(g,\lambda ) =\, \frac{1}{3}\, (4\pi)^{1-\frac{d}{2}}\Bigg[ \bigg\{
	d(d+1)\, \Phi_{d/2-1}^1(-2\lambda) \bigg\}_\text{dia}
	+ \bigg\{ -4d\, \Phi_{d/2-1}^1(0) \bigg\}_\text{ghost-dia} \nonumber\\
	&+ \bigg\{ -6d(d-1)\, \Phi_{d/2}^2(-2\lambda) \bigg\}_\text{para}
	+ \bigg\{-24\, \Phi_{d/2}^2(0) \bigg\}_\text{ghost-para} \,\Bigg] g
	+ \ord\left( g^2 \right).
\label{eqn:etaNexpansion}
\end{align}
Already at the level of (\ref{eqn:etaNexpansion}) an important observation can be made if
we take into account that all $\Phi$'s are strictly positive functions: \emph{The
graviton's paramagnetic contribution as well as both ghost terms (dia- and paramagnetic)
drive $\eta_N$ towards a negative value, while the graviton's diamagnetic part comes with
the opposite sign trying to make $\eta_N$ positive}.

Why is the sign of the anomalous dimension so crucial? This can be seen by considering
the definition $\eta_N \equiv k\p_k \ln G_k$: gravitational antiscreening, i.e.\ $G_k$
increases for decreasing RG scale $k$, amounts to $\eta_N<0$. Furthermore, as it is the
fundamental ingredient in the Asymptotic Safety scenario, our main interest will consist
in finding a non-Gaussian fixed point $(g_*,\lambda_*)$. By eq.\ (\ref{eqn:betag}), a
nontrivial fixed point ($g_*\neq 0$) requires that $\eta_N(g_*,\lambda_*) = 2-d$. Thus,
for any $d>2$, \emph{the occurrence of an NGFP is possible only if $\eta_N$ is negative}.

With regard to (\ref{eqn:etaNexpansion}) the fundamental requirement for an
asymptotically safe running Newton constant is that the graviton-diamagnetic effect is
weaker than the sum of the three other ones.

\subsection{Paramagnetic dominance: the basis for Asymptotic Safety}
\label{sec:ParaDomAS}
\begin{figure}[tp]
\begin{minipage}{.49\columnwidth}
	\flushleft
	\includegraphics[width=\columnwidth]{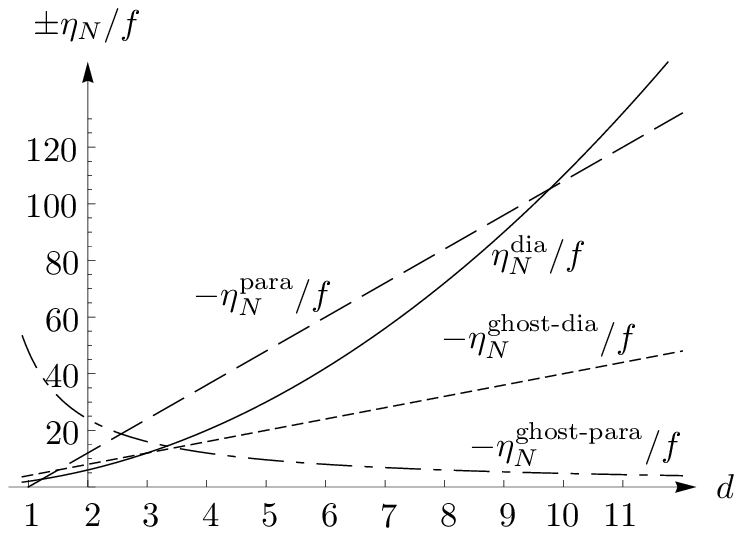}
\end{minipage}
\hfill
\begin{minipage}{.49\columnwidth}
	\flushright
	\includegraphics[width=\columnwidth]{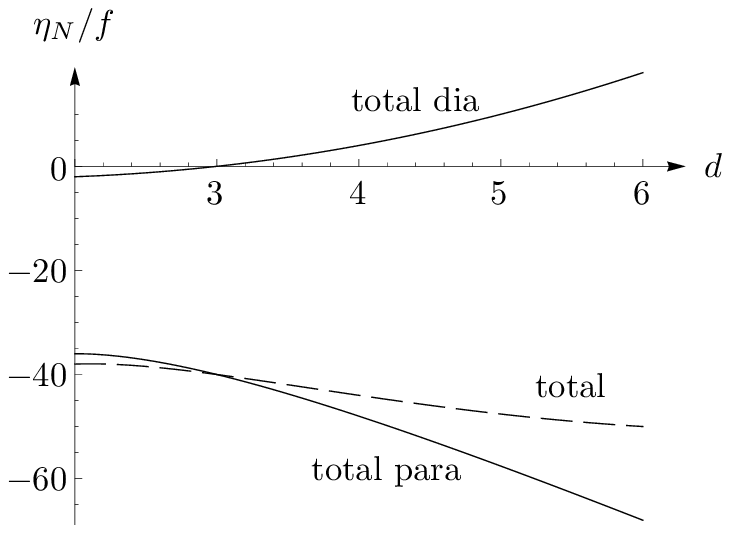}
\end{minipage}
\caption{Relative size of the various magnetic contributions to $\eta_N/f$, where 
		$f=\frac{1}{3\Gamma(d/2)}(4\pi)^{1-\frac{d}{2}}\,g \,$.
		On the left panel the absolute values of all four contributions (dia, para, ghost-dia
		and ghost-para) are shown. (Note that $\eta_N^\text{dia}$ has an opposite sign.)
		The right panel combines graviton-dia and ghost-dia as well as graviton-para and
		ghost-para parts. Here the total paramagnetic term overbalances the diamagnetic one,
		rendering the sum $\eta_N=\eta_N^\text{total dia}+\eta_N^\text{total para}$ negative
		(dashed line).}
	\label{fig:etaNtot}
\end{figure}

In order to compare the relative magnitude of the various magnetic contributions to
$\eta_N$ we may first employ a particular cutoff and demonstrate the universality of our
findings thereafter \cite{andreas1}. Here we choose the ``optimized'' shape function,
given by $\RN(z)=(1-z)\Theta(1-z)$. Then $\eta_N$ assumes the explicit form
\be
\begin{split}
\eta_N(g,\lambda) =\, \frac{1}{3}\,(4\pi)^{1-\frac{d}{2}}& \, \frac{1}{\Gamma(d/2)}\,
	\Bigg[ \bigg\{ \frac{d(d+1)}{1-2\lambda} \bigg\}_\text{dia}
	+ \bigg\{ -4d \bigg\}_\text{ghost-dia} \\
	+ &\bigg\{ -\frac{12(d-1)}{(1-2\lambda)^2} \bigg\}_\text{para}
	+ \bigg\{ -\frac{48}{d} \bigg\}_\text{ghost-para} \,
	\Bigg] g + \ord(g^2) \,.
\end{split}
\label{eqn:etaNopt}
\ee
The relative importance of the various terms in (\ref{eqn:etaNopt}) can be figured out
by setting $\lambda=0$ first, and considering $\lambda \neq 0$ subsequently.
Comparing the absolute values of the four curly brackets in (\ref{eqn:etaNopt}), for
$\lambda=0$, the most important result is that for any $d \lesssim 9.8$ there is indeed
always a negative contribution present which dominates the positive diamagnetic one. For
$2.6 \lesssim d \lesssim 9.8$ it is the graviton-paramagnetic part that provides the
largest contribution, while for $d \lesssim 2.6$ the ghost-paramagnetic effect is most
important, see left panel of figure \ref{fig:etaNtot}. Only for $d \gtrsim 14.4$ the
graviton-diamagnetic term would be large enough to win against the sum of the three other
ones and flip the sign of $\eta_N$. In $d=4$ for instance, we find the hierarchy
\be
\big\{|-36|\big\}_\text{para} > \big\{|+20|\big\}_\text{dia} >
	\big\{|-16|\big\}_\text{ghost-dia} > \big\{|-12|\big\}_\text{ghost-para} \,.
\ee
So the sign of $\eta_N$ is indeed determined by the three non-graviton-diamagnetic
contributions.

It is instructive to combine the graviton-para- and ghost-para-terms in a total
paramagnetic contribution, and similarly in the diamagnetic case. In this way we obtain
from (\ref{eqn:etaNopt}), at $\lambda=0$:
\be
\eta_N(g,0) =\, \frac{1}{3\Gamma(d/2)}\, (4\pi)^{1-\frac{d}{2}} \Big[ \,
	\big\{ d(d-3) \big\}_\text{total dia} - \big\{ 12(d-1)+48/d \big\}_\text{total para}\,
	\Big] g	+ \ord(g^2).
\label{eqn:etaNtot}
\ee
While the total paramagnetic part is always negative, we observe a sign change at $d=3$
in the total dia component. For $d<3$, the latter no longer counteracts the
paramagnetic interactions, but rather amplifies their effect of making $\eta_N$ negative.
This sign flip at $d=3$ is cutoff-independent, it holds for any choice of $\RN$.

The vanishing of the total diamagnetic contribution in 3 dimensions is closely related to
the fact that for $d=3$ there are no ``physical'' propagating degrees of freedom. (In
general there are $\frac{1}{2}\,d(d-3)$ of them which vanishes at $d=3$.) A detailed
explanation of this connection can be found in Ref.\ \refcite{andreas1}. Due to eq.\
(\ref{eqn:etaNtot}) the RG running of $g_k$ in $d=3$ is driven by the paramagnetic term
alone.

The relative total contributions to $\eta_N$ in (\ref{eqn:etaNtot}) are illustrated on
the right panel of figure \ref{fig:etaNtot}.
Here one clearly sees that, at least qualitatively, $\eta_N$ is determined only by the
total paramagnetic term, for all dimensions $d$ under consideration. In particular, the
negative sign arises only due to paramagnetism.

Finally we turn to the general case $\lambda \neq 0$. A careful analysis of the
$\beta$-functions shows that the fixed point value of the cosmological constant,
$\lambda_*$, is positive in basically all cases of interest \cite{andreas1}. So it is
sufficient to consider eq.\ (\ref{eqn:etaNopt}) with a positive $\lambda > 0$. The
paramagnetic contribution is already known to be dominant compared to the diamagnetic one
for $\lambda=0$. Going to larger values for $\lambda$ will even enhance this effect due
to the factor $(1-2\lambda)^{-2}$ in the paramagnetic part. Thus, also for general
$\lambda$, the crucial negative sign of $\eta_N$ in the fixed point regime stems from the
dominant paramagnetic terms.

\subsection[Phase portrait and NGFP in \texorpdfstring{$d=4$}{d = 4}]
	{Phase portrait and NGFP in $\bm{d=4}$}
\label{sec:FPQEG4d}

\begin{figure}[tp]
\begin{minipage}{0.49\textwidth}
	\centering
	\includegraphics[width=\columnwidth]{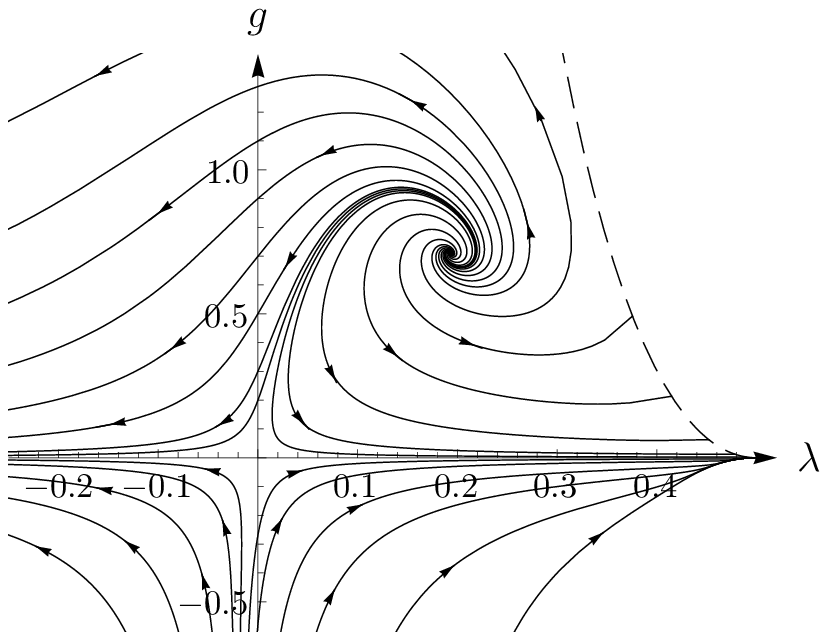}
\end{minipage}
\hfill
\begin{minipage}{0.49\textwidth}
	\centering
	\includegraphics[width=\columnwidth]{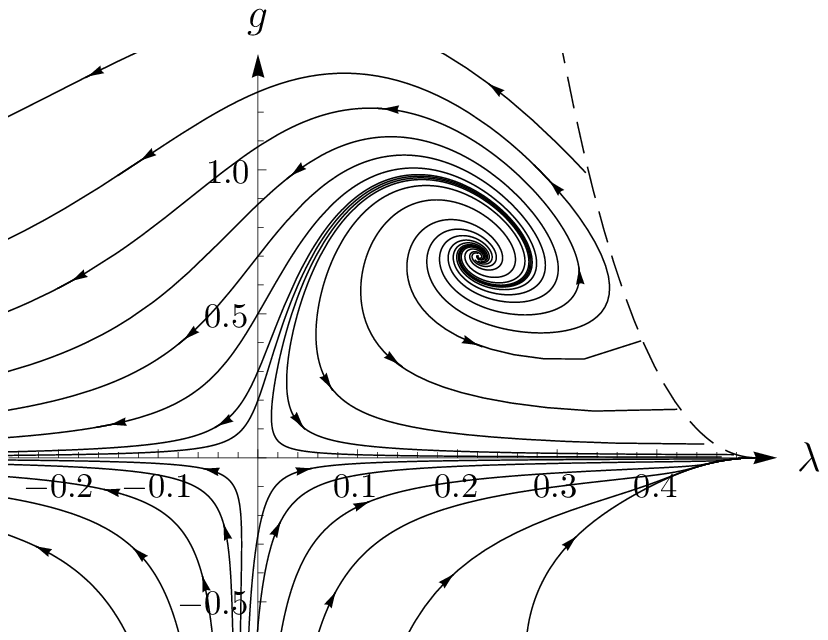}
\end{minipage}	
\caption{Standard phase portrait based on the full $\beta$-functions (left panel) and the
	approximation $\eta_N \approx g\, B_1(\lambda)$, right panel. In each diagram both dia-
	and paramagnetic contributions are retained.}
\label{fig:QEGfull}
\end{figure}

Specializing for 4 dimensions, we first recall the flow implied by the full
$\beta$-functions \cite{mr,frank1}, including all contributions to the anomalous
dimension. Then we show that restricting ourselves to the linear approximation (in $g$)
of $\eta_N$ leads to essentially the same result. Afterwards we perform the same
computation, but this time we consider only paramagnetic terms in $\eta_N$. Finally, we
repeat the latter step using diamagnetic contributions only.

\noindent
\textbf{(i)} We start with the RG equations (\ref{eqn:betalambda}) and (\ref{eqn:betag})
together with the full anomalous dimension (\ref{eqn:etaN}), employing the optimized
cutoff. The resulting phase portrait, obtained by a numerical evaluation, is the well
known one \cite{frank1}; it is depicted on the left panel in figure \ref{fig:QEGfull}.
One finds a Gaussian fixed point in the origin, but also a UV attractive non-Gaussian
fixed point. The dashed curve restricts the domain of the $g$-$\lambda$ theory space
since there the $\beta$-functions diverge. To the left of this boundary all points with
positive Newton's constant are ``pulled'' into the NGFP for $k \rightarrow \infty$. (The
arrows in all flow diagrams point from the UV to the IR.)
\medskip

\noindent
\textbf{(ii)} To show that the denominator in $\eta_N = g\, B_1(\lambda)\big/\big(1- g\,
B_2(\lambda)\big)$ leads only to qualitatively inessential modifications we solve the RG
equations with the approximate anomalous dimension obtained in leading order of the
$g$-expansion, $\eta_N \approx g\, B_1(\lambda)$, as given in (\ref{eqn:etaNopt}),
retaining both dia- and paramagnetic terms. The resulting phase portrait, shown on the
right panel of figure \ref{fig:QEGfull}, is basically indistinguishable from the exact
one on the left. Therefore, we may continue with the approximation
$\eta_N \approx g\, B_1(\lambda)$.
\medskip

\begin{figure}[tp]
\begin{minipage}{.49\columnwidth}
	\includegraphics[width=\columnwidth]{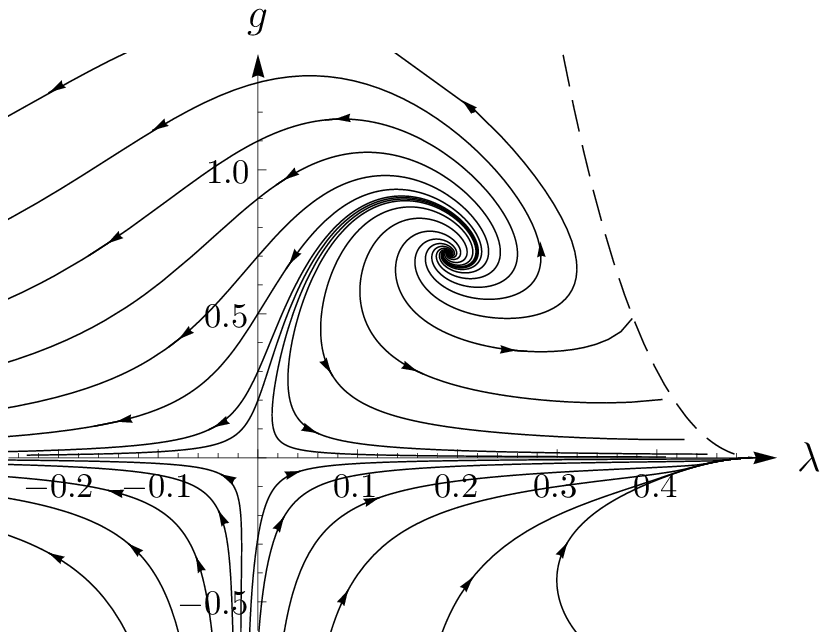}
\end{minipage}
\hfill
\begin{minipage}{.49\columnwidth}
	\includegraphics[width=\columnwidth]{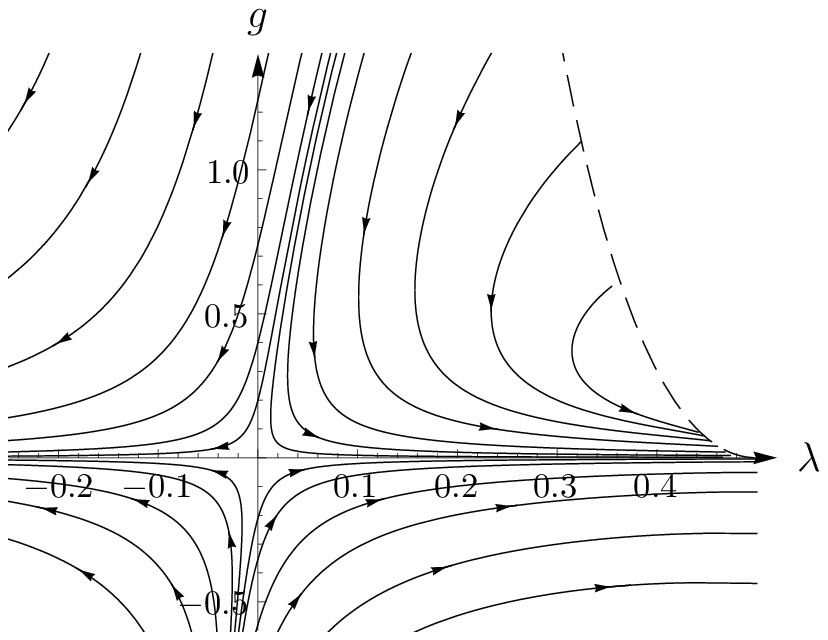}
\end{minipage}
\caption{Flow diagram obtained from the \emph{total paramagnetic contributions} to
	$\eta_N$ alone (left panel), and taking into account the \emph{total diamagnetic terms}
	in $\eta_N$ only (right panel).}
\label{fig:QEGparadia}
\end{figure}

\noindent
\textbf{(iii)} Next we use eq.\ (\ref{eqn:etaNopt}) again, but take into account the
\emph{total paramagnetic contributions only}. Thus $\eta_N$ assumes the simple form
\be
\eta_N^\text{total para}(g,\lambda) = -\frac{1}{\pi} \left[ \frac{3}{(1-2\lambda)^2} + 1
	\right]	g + \ord(g^2) \,,
\label{eqn:etaPara}
\ee
where the first term inside the brackets of (\ref{eqn:etaPara}) is due to the gravitons,
while the ``+1'' stems from the ghosts. We insert this expression into the
$\beta$-functions of $g$ and $\lambda$, and again obtain the flow by a numerical
computation. The left panel of figure \ref{fig:QEGparadia} displays the resulting phase
portrait. The similarity of this diagram to the phase portrait based on the full
$\beta$-functions in figure \ref{fig:QEGfull} is truly impressive. All qualitative
features of the flow are incorporated already in the total paramagnetic terms alone. In
particular we find an NGFP with two UV-attractive eigendirections. Even the values of the
fixed point coordinates and critical exponents do not change significantly
\cite{andreas1}. This shows that paramagnetism is at the heart of Asymptotic
Safety.
\medskip

\noindent
\textbf{(iv)} At last we perform the same steps as in (iii), but keep only the
\emph{total diamagnetic contributions} to $\eta_N$ in (\ref{eqn:etaNopt}), such that it
is given by
\be
\eta_N^\text{total dia}(g,\lambda) =\, \frac{1}{3\pi} \left[ \frac{5}{1-2\lambda} - 4
	\right] g + \ord(g^2) \,,
\ee
where the ``-4'' comes from the ghosts. This anomalous dimension leads to the flow
diagram depicted in the right panel of figure \ref{fig:QEGparadia}. The structure of the
flow is quite different now, in particular the NGFP has disappeared.
This illustrates that the total diamagnetic term contributes to $\eta_N$ with the
``wrong'' sign and rather counteracts the emergence of an NGFP. Hence, diamagnetic
effects work against Asymptotic Safety. Note that the culprit is alone the diamagnetism
of the graviton; the ghosts make a negative contribution to $\eta_N^\text{total dia}$ and
actually favor an NGFP.

The above results can be proven \cite{andreas1} to be independent of the cutoff chosen.
Thus we can conclude that \emph{the formation of an NGFP in the RG flow of QEG is a
universal result of the paramagnetic interaction of the metric fluctuations with their
background}.

\subsection[The \texorpdfstring{$\beta$}{b}-function of \texorpdfstring{$g$}{g} in
	\texorpdfstring{$2+\epsilon$}{2 + e} dimensions]{The $\bm{\beta}$-function of
	$\bm{g}$ in \bm{$2+\epsilon$} dimensions}
\label{sec:FPQEG2d}
As Newton's constant becomes dimensionless in two dimensions the RG flow of the
gravitational average action shows a certain degree of universality if $d=2+\epsilon$ for
$\epsilon$ small \cite{mr}: the relevant threshold functions at leading order in
$\epsilon$ read $\Phi_1^2(\lambda_k)=\Phi_1^2(0)+\ord(\epsilon)$ and
$\Phi_0^1(\lambda_k)=\Phi_0^1(0)+ \ord(\epsilon)$ where $\Phi_1^2(0)=1$ and
$\Phi_0^1(0)=1$ for any cutoff shape function $\RN$. As a result, the anomalous dimension
can be written as $\eta_N=-b\, g+\ord(g^2)$ where the universal coefficient $b$, in its
decomposed form, follows from (\ref{eqn:etaNexpansion}): $b = \frac{1}{3}\, \big[
\{-6\}_\text{dia} + \{8\}_\text{ghost-dia}+ \{12\}_\text{para} +
\{24\}_\text{ghost-para} \big]$, or
\be
b = \frac{2}{3}\, \Big[ \{1\}_\text{total dia} + \{18\}_\text{total para} \Big]
	= \frac{38}{3} \,.
\label{eqn:bcoefftot}
\ee
Thus the crucial number $b$ is positive -- and the anomalous dimension negative therefore
-- not only thanks to the large ``para'' contribution but also because of the smaller,
but positive diamagnetic one. This is exactly as it should be since we know that below
$d=3$ the diamagnetic interaction drives $\eta_N$ in the same direction as the
paramagnetic, see figure \ref{fig:etaNtot}.

In the literature \cite{GKTCD,camp38} there has been a
considerable amount of confusion about the correct value of $b$. The two classes of
disagreeing results were quoted already in Ref.\ \refcite{Weinberg} by Weinberg.
According to Refs.\ \refcite{camp38},
\be
b = \frac{38}{3} \,.
\label{eqn:bcamp38}
\ee
The authors of Refs.\ \refcite{GKTCD} find instead $b = 2/3$.

Comparing the results of the two camps to our answer obtained by means of the effective
average action, eq.\ (\ref{eqn:bcoefftot}), we observe that \emph{the first candidate,
the coefficient in} (\ref{eqn:bcamp38}), \emph{amounts to the full, i.e.\ dia- plus
paramagnetic contribution, while the second,} $b = 2/3$, \emph{consists of
the diamagnetic one only.} Looking at the details of their respective derivations one can
see that the different treatment of the paramagnetic piece is indeed the source of the
disagreement.

The result $b = 2/3$ found in Refs.\ \refcite{GKTCD} is by no means wrong, but
rather amounts to a different definition of the running Newton constant, namely via the
coefficient of the Gibbons-Hawking surface term. And indeed, when we use the FRGE to
compute the running of this boundary Newton constant the result we find agrees with
Refs.\ \refcite{GKTCD}: Let $g^\p$ denote Newton's constant on the boundary and $\eta_N^\p$
its anomalous dimension. Then $\eta_N^\p$, to leading order in the ``bulk'' constant $g$,
is given by \cite{boundary}
\be
\eta_N^\p(g^\p,g,\lambda) = \frac{1}{3}(4\pi)^{1-\frac{d}{2}}\left[ d(d+1)\,
	\Phi_{d/2-1}^1(-2\lambda) -4d\,\Phi_{d/2-1}^1(0) \right]g^\p \,.
\label{eqn:etaboundary}
\ee
Going through the derivation of (\ref{eqn:etaboundary}) it is easy to see that in leading
order \emph{the surface anomalous dimension $\eta_N^\p$ is of entirely diamagnetic
origin}. In $d=2+\epsilon$ it becomes $\eta_N^\p = - b^\p\,g^\p + \cdots$ with $b^\p =
2/3$, coinciding precisely with Refs.\ \refcite{GKTCD}. This confirms that the authors
advocating $b=2/3$ actually computed the anomalous dimension of the boundary
Newton constant, while those who obtained (\ref{eqn:bcamp38}) focused on the bulk
quantity. Thus, in a way, both camps are right, but their respective results, $\eta_N^\p$
and $\eta_N$, are unavoidably different as a consequence of the paramagnetic interaction.

\section{QEG spacetimes as a polarizable medium}
\label{sec:QEGvacuum}
The previous sections dealt with the predominance of \emph{paramagnetic interactions}
over diamagnetic ones in determining certain gross features of the RG flow in QEG. A
priori this fact has nothing to do with the interpretation of the quantum field theory
vacuum as a \emph{paramagnetic medium}. While the previous sections all dealt with the
dominance of the paramagnetic interaction term over the diamagnetic one, we now turn to
the question of how the QEG vacuum responds to external fields. As we will argue, the
spacetimes of QEG are analogous to the vacuum state of Yang-Mills theory since they can
be seen as a polarizable medium with a ``paramagnetic'' response to external
perturbations. Rather than introducing tensorial susceptibilities for a general
gravitational field we restrict the expectation value of the metric to the form of the
lowest post-Newtonian order. This will display the analogy of QEG to QED or Yang-Mills
theory most clearly. We consider metrics
\be
g_\mn dx^\mu dx^\nu = -(1 + 2\,\Phi_\text{grav})\, dt^2 +2\,\bm{\zeta}\cdot d\mathbf{x}\,
	dt + (1 - 2\,\Phi_\text{grav})\, d\mathbf{x}^2 \,,
\label{eqn:metricpostN}
\ee
with $\Phi_\text{grav}$ and $\bm{\zeta}$ time independent gravitational scalar and vector
potentials, respectively, satisfying the harmonic coordinate condition,
$\bm{\nabla}\cdot\bm{\zeta} = 0$. Leaving the cosmological constant aside, we employ the
Lorentzian version \cite{Lor-EAA} of the effective average action $\Gamma_k^\text{Lor}[g]
= \frac{1}{16\pi G_k} \int \text{d}^4 x \sqrt{-g}\, R[g]$. Inserting the metric
(\ref{eqn:metricpostN}) and retaining at most quadratic terms in $\Phi_\text{grav}$ and
$\bm{\zeta}$ we find
\be
\Gamma_k^\text{Lor}[g] = - \frac{1}{4\pi} \int \text{d}^4 x\,\frac{1}{2G_k}\,
	\big( \mathbf{E}^2_\text{grav} - \mathbf{B}^2_\text{grav} \big) \,.
\label{eqn:GammaLor}
\ee
Here we encounter the acceleration $\mathbf{E}_\text{grav} \equiv - \bm{\nabla}
\Phi_\text{grav}$ and the angular velocity of the local inertial frames,
$\mathbf{B}_\text{grav} \equiv - \frac{1}{2}\bm{\nabla} \times \bm{\zeta}$. They are the
gravitational analogs of the electromagnetic, or Yang-Mills, $\mathbf{E}$ and
$\mathbf{B}$ fields, respectively. It turns out natural to rewrite eq.\
(\ref{eqn:GammaLor}) in the following fashion:
\be
\Gamma_k^\text{Lor}[g] = -\frac{1}{4\pi} \int \text{d}^4 x\, \frac{1}{2G_\Lambda}
	\left( \varepsilon_k^\text{grav} \mathbf{E}^2_\text{grav} -
	\frac{1}{\mu_k^\text{grav}}\, \mathbf{B}^2_\text{grav} \right) \,.
\label{eqn:GammaLorNew}
\ee
Here $G_\Lambda$ is Newton's constant at some fixed UV scale $k = \Lambda$. The
$k$-dependence of the action is carried by the ``gravi-dielectric constant''
$\varepsilon_k^\text{grav}$ and the ``gravimagnetic permeability'' $\mu_k^\text{grav}$,
defined by
\be
\varepsilon_k^\text{grav} = \frac{1}{\mu_k^\text{grav}} = \frac{G_\Lambda}{G_k} \;.
\ee
For the simplified trajectory $G_k^{-1}=G_0^{-1}+k^2/g_*$, for example, we obtain
the relation
$\varepsilon_k^\text{grav} = 1/\mu_k^\text{grav} = (g_*+G_0 k^2)/(g_*+G_0\Lambda^2)$.
The analogy between (\ref{eqn:GammaLorNew}) and its gauge theory counterpart,
$\frac{1}{2}\,g_\Lambda^{-2}\left( \varepsilon_k \mathbf{E}^2 - \mu_k^{-1}
\mathbf{B}^2 \right)$, is indeed quite striking.

In order to understand the physics contents of (\ref{eqn:GammaLorNew}) let us ask how
$\varepsilon_k^\text{grav}$ and $\mu_k^\text{grav}$ evolve along an RG trajectory.
Starting at the UV cutoff $k = \Lambda$ we have $\varepsilon_\Lambda^\text{grav} =
\mu_\Lambda^\text{grav} = 1$ initially. Then, lowering $k$, the RG flow is such that
$G_k$ is the larger the smaller is $k$. Hence we see that integrating out the metric
fluctuations in the momentum interval $[k,\Lambda]$ gives rise to a gravi-dielectric
constant (gravimagnetic permeability) smaller (larger) than unity:
\be
\varepsilon_k^\text{grav} \leq 1 \,, \quad \mu_k^\text{grav} \geq 1 \qquad \text{for }\,
	k \leq \Lambda.
\ee
In this sense, \emph{the behavior of the QEG vacuum is analogous to that of Yang-Mills
theory: $\varepsilon_k^\mathrm{grav} < 1$ implies that external charges (masses) are
antiscreened, and $\mu_k^\mathrm{grav} > 1$ indicates the paramagnetic response to
external gravimagnetic fields}.

\section{Summary and conclusion}
\label{sec:conclusion}
In this article we identified and described a general physical mechanism which seems to
underly the nonperturbative renormalizability, or Asymptotic Safety, of Quantum Einstein
Gravity according to all investigations available to date. Our discussion started from
the observation that the RG flow of QEG is driven by the quantum fluctuations of the
metric, and that those are governed by an inverse propagator of the general form
$-D_\cA^2+\mU(F_\cA)$. Related to the different interactions the fluctuations can have
with the background we introduced a generalized notion of dia- and paramagnetism, and
then disentangled the dia- from the para-type contributions to the RG flow, in particular
to the anomalous dimension of Newton's constant, $\eta_N$. The negative sign of $\eta_N$
which is crucial for gravitational antiscreening and Asymptotic Safety was found to be
due to the predominantly paramagnetic interaction of the gravitons with external
gravitational fields. Those interactions are sufficient by themselves to trigger the
formation of a non-Gaussian RG fixed point. On the other hand, the diamagnetic
interaction would not lead to such a fixed point on its own, and, in fact, in $d > 3$
dimensions it counteracts antiscreening and Asymptotic Safety. Thus \emph{the NGFP owes
its existence to the paramagnetic dominance}.

In the familiar quantum field theories, such as QED and QCD, one of the most interesting
tasks, which often is also essential from the practical point of view, consists in
determining the properties of its vacuum state, e.g.\ the response of quantum
fluctuations to external fields. In the case of quantum gravity we were led to the
following intuitive picture of a QEG ``vacuum'' state, a spacetime represented by a
self-consistent solution $\bar{g}_\mn$ to the effective field equations for instance.

The dominant paramagnetic coupling of the metric fluctuations $h_\mn$ to their
``condensate'', that is, the background $\bar{g}_\mn$, has the form $\int h(x) \bar{U}(x)
h(x)$ which is analogous to $\int \bar{\psi} (\bm{\sigma} \cdot \mathbf{B}) \psi$ for
magnetic systems. It contains no derivatives of $h_\mn$, i.e.\ it is \emph{ultralocal},
and the interaction energy it gives rise to depends only on the spin orientation of
$h(x)$ relative to $\bar{U}(x)$ at each spacetime point $x$ individually. So the
essential physical effects in the fixed point regime are due to \emph{fluctuations which
do not correlate different spacetime points}. To the extent the orbital motion effects
caused by $\int h \bar{D}^2 h$ can be neglected, different spacetime points decouple
completely.

Thus, if one wants to invoke a magnetic analogy again, the QEG vacuum should be
visualized as a statistical spin system which consists of magnetic moments sitting at
fixed lattice points and interacting with their mean field, rather than as a gas of
itinerant electrons.

\end{document}